\documentclass[conference]{IEEEtran}
\IEEEoverridecommandlockouts
\usepackage{cite}
\usepackage{amsmath,amssymb,amsfonts}
\usepackage{algorithmic}
\usepackage{graphicx}
\usepackage{textcomp}
\usepackage{xcolor}
\usepackage{tikz}
\usetikzlibrary{positioning, arrows.meta, calc}

\usepackage{braket}
\usepackage{pgfplots}
\pgfplotsset{compat=1.18}
\usepgfplotslibrary{groupplots}
\usepackage{enumitem}

\def\BibTeX{{\rm B\kern-.05em{\sc i\kern-.025em b}\kern-.08em
    T\kern-.1667em\lower.7ex\hbox{E}\kern-.125emX}}
    
\begin{document}

\title{Hardware Robustness of Sample-Based Quantum Diagonalization}

\author{\IEEEauthorblockN{1\textsuperscript{st} Ahatesham Bhuiyan}
\IEEEauthorblockA{\textit{University of Central Florida}\\
Orlando, FL, USA \\
ahatesham.bhuiyan@ucf.edu}
\and
\IEEEauthorblockN{2\textsuperscript{nd} Cheng Chu}
\IEEEauthorblockA{\textit{North Carolina State University}\\
Raleigh, NC, USA \\
cchu8@ncsu.edu}
\and
\IEEEauthorblockN{3\textsuperscript{rd} Qian Lou}
\IEEEauthorblockA{\textit{University of Central Florida}\\
Orlando, FL, USA \\
qian.lou@ucf.edu}
\and
\IEEEauthorblockN{4\textsuperscript{th} Mengxin Zheng}
\IEEEauthorblockA{\textit{University of Central Florida}\\
Orlando, FL, USA \\
mengxin.zheng@ucf.edu}
}

% \and
% \and
% \and
% \and
% \and

\maketitle

\begin{abstract}
% Sample-based Quantum Diagonalization (SQD) is a hybrid quantum-classical method that replaces variational optimization with a self-consistent recovery loop over QPU samples. SQD is generally considered to be robust to noisy quantum samples and imperfect classical inputs, yet how this robustness behaves on real hardware across the deployment choices a practitioner makes has not been systematically characterized. As a result, shot budgets, qubit layouts, noise mitigation strategies, and the classical coupled-cluster singles and doubles (CCSD) amplitudes that anchor the ansatz are typically selected without empirical guidance. We present a robustness analysis of SQD on IBM Heron hardware along these three dimensions. Structured perturbations of the CCSD amplitudes, including complete zeroing, shift the recovered energy only modestly from the clean baseline. Differences across qubit layouts and noise mitigation settings are large at the first recovery iteration but narrow to a small band within a few iterations. Post-recovery accuracy saturates at moderate shot budgets, with very large shot budgets yielding slightly worse recovered energies than smaller ones, a behavior we attribute to how the recovery loop selects its working set of bitstrings, which limits the information that additional shots can contribute. Together, these findings show where SQD's recovery loop confers genuine robustness against deployment choices and where its limits lie, providing a hardware-grounded basis for understanding when SQD's resilience can be relied upon.

Sample-based Quantum Diagonalization (SQD) is a hybrid quantum-classical method that replaces variational optimization with a self-consistent recovery loop over QPU samples. Although SQD is considered robust to noisy samples and imperfect classical inputs, its robustness across practical deployment choices has not been systematically analyzed. As a result, shot budgets, qubit layouts, noise mitigation strategies, and the coupled-cluster singles and doubles (CCSD) amplitudes that initialize the ansatz are often chosen without clear empirical guidance. We analyze SQD robustness on IBM Heron hardware across these dimensions. Structured CCSD-amplitude perturbations, including complete zeroing, produce only modest energy shifts from the clean baseline. Differences across layouts and noise-mitigation settings are large in the first recovery iteration but narrow within a few iterations. Accuracy saturates at moderate shot budgets, while very large budgets slightly worsen recovered energies, likely because working-set selection limits the value of additional samples. These results identify where SQD provides genuine deployment robustness and where its limits remain.
\end{abstract}


\section{Introduction}
\label{sec:introduction}

Quantum chemistry is a leading near-term application of quantum computing, but variational quantum eigensolver (VQE) methods face well-known challenges from barren plateaus, large measurement overheads, and optimizer instability under shot noise~\cite{b1,b2,b3}. Sample-based Quantum Diagonalization (SQD) offers a different route: rather than optimizing circuit parameters on the QPU, it samples bitstrings from a fixed local unitary cluster Jastrow (LUCJ) circuit, applies a self-consistent configuration recovery loop, and classically diagonalizes the Hamiltonian in the recovered subspace~\cite{b4,b5}. By removing the variational optimizer, SQD has been demonstrated on active spaces of up to $36$ orbitals using $77$ qubits~\cite{b4}.

Although SQD is considered robust to noisy samples and imperfect classical inputs, its behavior under practical deployment choices has not been systematically analyzed. In current workflows, practitioners must choose coupled-cluster singles and doubles (CCSD)~\cite{b6,b9} amplitudes for LUCJ initialization, qubit layouts and noise-mitigation settings for hardware execution, and QPU shot budgets for sampling. These choices affect cost and hardware performance, yet their impact on the recovered energy remains unclear.

\definecolor{cgrayfill}{HTML}{F1EFE8}
\definecolor{cgraystroke}{HTML}{5F5E5A}
\definecolor{cpurplefill}{HTML}{EEEDFE}
\definecolor{cpurplestroke}{HTML}{534AB7}
\definecolor{cpurpletext}{HTML}{3C3489}
\definecolor{ccoralfill}{HTML}{FAECE7}
\definecolor{ccoralstroke}{HTML}{993C1D}
\definecolor{ccoraltext}{HTML}{712B13}
\definecolor{cbluefill}{HTML}{E6F1FB}
\definecolor{cbluestroke}{HTML}{185FA5}
\definecolor{cbluetext}{HTML}{0C447C}
\definecolor{camberfill}{HTML}{FAEEDA}
\definecolor{camberstroke}{HTML}{854F0B}
\definecolor{cambertext}{HTML}{633806}
\definecolor{ctealstroke}{HTML}{0F6E56}

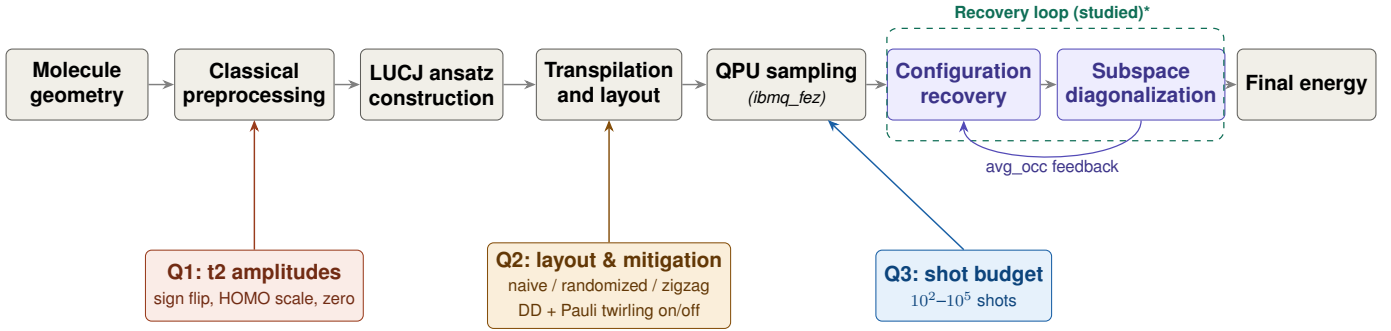
\begin{figure*}[t]
\centering
\resizebox{\textwidth}{!}{%
\begin{tikzpicture}[
  font=\sffamily,
  >={Stealth[length=2mm, width=1.6mm]},
  stage/.style={
    draw=cgraystroke, fill=cgrayfill,
    rounded corners=3pt,
    minimum width=2.4cm, minimum height=1.2cm,
    line width=0.5pt, align=center, inner sep=4pt
  },
  loopstage/.style={
    draw=cpurplestroke, fill=cpurplefill,
    rounded corners=3pt,
    minimum width=2.4cm, minimum height=1.2cm,
    line width=0.5pt, align=center, inner sep=4pt,
    text=cpurpletext
  },
  endstage/.style={
    draw=cgraystroke, fill=cgrayfill,
    rounded corners=3pt,
    minimum width=2.0cm, minimum height=1.2cm,
    line width=0.5pt, align=center, inner sep=4pt
  },
  probet2/.style={
    draw=ccoralstroke, fill=ccoralfill,
    rounded corners=3pt,
    minimum width=2.6cm, minimum height=1.2cm,
    line width=0.5pt, align=center, inner sep=4pt, text=ccoraltext
  },
  probelayout/.style={
    draw=camberstroke, fill=camberfill,
    rounded corners=3pt,
    minimum width=3.0cm, minimum height=1.2cm,
    line width=0.5pt, align=center, inner sep=4pt, text=cambertext
  },
  probeshot/.style={
    draw=cbluestroke, fill=cbluefill,
    rounded corners=3pt,
    minimum width=2.4cm, minimum height=1.2cm,
    line width=0.5pt, align=center, inner sep=4pt, text=cbluetext
  },
  pipearrow/.style={->, line width=0.7pt, gray}
]

\node[stage] (geom) at (0, 0)
  {\textbf{Molecule}\\\textbf{geometry}};
\node[stage] (classical) at (3.0, 0)
  {\textbf{Classical}\\\textbf{preprocessing}};
\node[stage] (ansatz) at (6.0, 0)
  {\textbf{LUCJ ansatz}\\\textbf{construction}};
\node[stage] (trans) at (9.0, 0)
  {\textbf{Transpilation}\\\textbf{and layout}};
\node[stage] (qpu) at (12.0, 0)
  {\textbf{QPU sampling}\\\footnotesize\textit{(ibmq\_fez)}};
\node[loopstage] (score) at (15.0, 0)
  {\textbf{Configuration}\\\textbf{recovery}};
\node[loopstage] (dav) at (18.0, 0)
  {\textbf{Subspace}\\\textbf{diagonalization}};
\node[endstage] (energy) at (20.8, 0)
  {\textbf{Final energy}};

\draw[ctealstroke, dashed, line width=0.6pt, rounded corners=5pt]
  (13.7, -0.95) rectangle (19.4, 0.95);
\node[font=\sffamily\footnotesize\bfseries, text=ctealstroke]
  at (16.5, 1.20) {Recovery loop (studied)*};

\draw[pipearrow] (geom.east)      -- (classical.west);
\draw[pipearrow] (classical.east) -- (ansatz.west);
\draw[pipearrow] (ansatz.east)    -- (trans.west);
\draw[pipearrow] (trans.east)     -- (qpu.west);
\draw[pipearrow] (qpu.east)       -- (score.west);
\draw[pipearrow] (score.east)     -- (dav.west);
\draw[pipearrow] (dav.east)       -- (energy.west);

\draw[->, line width=0.6pt, cpurplestroke]
  (dav.south) .. controls +(0,-0.8) and +(0,-0.8) .. (score.south);
\node[font=\sffamily\footnotesize, text=cpurpletext]
  at (16.5, -1.40) {avg\_occ feedback};

\node[probet2] (t2) at (3.0, -3.4)
  {\textbf{Q1: t2 amplitudes}\\\footnotesize sign flip, HOMO scale, zero};
\node[probelayout] (layout) at (9.0, -3.4)
  {\textbf{Q2: layout \& mitigation}\\\footnotesize naive / randomized / zigzag\\\footnotesize DD + Pauli twirling on/off};
\node[probeshot] (shot) at (15.0, -3.4)
  {\textbf{Q3: shot budget}\\\footnotesize $10^2$--$10^5$ shots};

\draw[->, line width=0.7pt, ccoralstroke]
  (t2.north) -- (3.0, -0.6);
\draw[->, line width=0.7pt, camberstroke]
  (layout.north) -- (9.0, -0.6);
\draw[->, line width=0.7pt, cbluestroke]
  (shot.north) -- (12.7, -0.6);

\node[font=\sffamily\footnotesize\itshape, text=gray, anchor=west]
  at (-1.2, -4.5) {* iterated until convergence with $K$ parallel batches};

\end{tikzpicture}%
}
\caption{SQD workflow and the three robustness probes (Q1--Q3) analyzed in this work.}
\label{fig:pipeline}
\end{figure*}

We present a hardware case study of SQD robustness on IBM's Heron-r2 processor (\texttt{ibmq\_fez}) across three pipeline-aligned probes:
\begin{itemize}[leftmargin=*, itemsep=1pt, topsep=2pt]
    \item \textbf{Q1:} We perturb the CCSD $t_2$ amplitudes used to initialize the LUCJ ansatz.
    \item \textbf{Q2:} We compare qubit-layout and noise-mitigation choices.
    \item \textbf{Q3:} We sweep QPU shot budgets to evaluate post-recovery accuracy.
\end{itemize}
At each recovery iteration, SQD filters or repairs bitstrings using particle-number constraints and average orbital occupancies, selects a working set of recovered configurations, and diagonalizes the Hamiltonian in that subspace. This makes the final energy depend on whether the recovered subspace contains important determinants, not only on the quality of individual raw samples.

Across the tested molecules, perturbations, and hardware runs, our findings show that the recovery loop absorbs many deployment differences:
\begin{itemize}[leftmargin=*, itemsep=1pt, topsep=2pt]
    \item \textbf{Q1:} Structured CCSD-amplitude changes, including complete zeroing, produce only modest shifts from the clean baseline.
    \item \textbf{Q2:} Layout and mitigation choices cause large differences in the first recovery iteration, but these differences narrow after a few iterations.
    \item \textbf{Q3:} Accuracy saturates at moderate shot budgets, with very large budgets sometimes producing slightly worse recovered energies.
\end{itemize}
These findings provide hardware-grounded evidence for where SQD offers practical robustness and where deployment choices still require care.

\section{SQD Pipeline}
\label{sec:background}

Fig.~\ref{fig:pipeline} summarizes the SQD data flow and the three deployment probes studied in this work. SQD is a hybrid quantum-classical method for estimating molecular ground-state energies without variational optimization on the QPU~\cite{b4,b10}. The quantum stage prepares a fixed local unitary cluster Jastrow (LUCJ) ansatz, a hardware-efficient fermionic circuit designed to represent electron correlation with relatively low depth on heavy-hex devices~\cite{b5}. Its parameters are initialized from coupled-cluster singles and doubles (CCSD), a classical quantum-chemistry method that approximates electron correlation by including single- and double-electron excitations from a Hartree--Fock reference state~\cite{b5,b11}. Measuring the LUCJ circuit produces computational-basis bitstrings; in the fermionic encoding used for electronic structure, each bitstring represents an occupation pattern over spin orbitals and corresponds to a candidate Slater determinant.

The classical stage converts these noisy candidate determinants into a diagonalization subspace. SQD first checks whether each measured bitstring satisfies physical constraints such as the required number of spin-up and spin-down electrons. Invalid strings are either discarded or repaired by flipping occupations according to average orbital occupancies estimated from the current sample set. The recovered strings are grouped into batches, and the molecular Hamiltonian is diagonalized within the span of each batch. The lowest-energy batch updates the orbital occupancies used in the next recovery iteration, forming a self-consistent loop~\cite{b10}.

\section{Robustness Analysis Approach}
\label{sec:approach}

We define three pipeline-aligned probes of the SQD recovery loop, shown in Fig.~\ref{fig:pipeline}. Each probe varies one practitioner-controlled input while the remaining settings are held at the defaults reported in Section~\ref{sec:experiments}.

\subsection{Q1: Classical-Input Perturbations}
\label{subsec:probe_t2}

Q1 perturbs the CCSD amplitudes used to initialize the LUCJ ansatz, modeling upstream corruption or reuse of imperfect classical inputs. CCSD produces $t_1$ amplitudes for single-electron excitations and $t_2$ amplitudes for two-electron excitations~\cite{b11}. We focus on $t_2$ because double excitations capture a large share of electron correlation and directly shape the LUCJ initialization. Here, $t_2[i,j,a,b]$ denotes the amplitude for exciting two electrons from occupied orbitals $i,j$ into virtual orbitals $a,b$. We define three structured perturbations $\mathcal{P}_i: t_2 \mapsto t_2'$:
\begin{align}
\mathcal{P}_1 \text{ (sign flip):}\quad &
  t_2'[i,j,a,b] = -\, t_2[i,j,a,b], \\
\mathcal{P}_2 \text{ (HOMO scaling):}\quad &
  t_2'[i,j,a,b] = 2\pi \cdot t_2[i,j,a,b],\; i \in \mathcal{H}, \\
\mathcal{P}_3 \text{ (zeroing):}\quad &
  t_2'[i,j,a,b] = 0.
\end{align}
Here, $\mathcal{H}$ denotes the occupied-orbital index set corresponding to the highest occupied molecular orbital (HOMO), and the $2\pi$ factor is chosen as a large nontrivial multiplier that strongly distorts HOMO-associated amplitudes while preserving their tensor support. These cases test complementary failures: global sign corruption, targeted distortion of frontier-orbital excitations, and complete removal of double-excitation correlation information.

\subsection{Q2: Layout and Mitigation Choices}
\label{subsec:probe_layout}

Q2 varies two hardware-execution choices for the LUCJ circuit: qubit layout and noise mitigation. Qubit layout maps logical qubits in the circuit to physical qubits on \texttt{ibmq\_fez}; because two-qubit gates must follow the device's heavy-hex connectivity, poor layouts can increase routing overhead and decoherence~\cite{b12,b13}. We compare three layouts: a \emph{naive} contiguous mapping, a \emph{randomized} mapping as an unstructured baseline, and a \emph{zigzag} mapping that places spin-$\alpha$ and spin-$\beta$ orbitals on neighboring heavy-hex rows with bridge qubits between them, following the LUCJ hardware-aware design~\cite{b5}. Noise mitigation refers to execution-time techniques that reduce hardware noise without changing the target circuit. For each layout, we run with mitigation enabled or disabled; enabled mitigation applies dynamical decoupling during idle periods~\cite{b14} and Pauli twirling to randomize coherent two-qubit gate errors~\cite{b15}, yielding six layout-mitigation configurations.

\subsection{Q3: Shot-Budget Scaling}
\label{subsec:probe_shots}
Q3 varies the QPU shot count $N_{\text{shots}}$ used to sample the LUCJ circuit. Because SQD diagonalizes a fixed-size recovered subspace rather than directly averaging all raw samples, additional shots may eventually stop improving the final energy. We therefore sweep $N_{\text{shots}}$ across orders of magnitude to identify whether post-recovery accuracy saturates.

\section{Experimental Setup}
\label{sec:experiments}

We evaluate SQD on three molecular active spaces: BeH\textsubscript{2}, a predominantly single-reference system; H\textsubscript{2}O, a closed-shell molecule with single-bond character; and N\textsubscript{2}, a more strongly correlated triple-bond system. BeH\textsubscript{2} and H\textsubscript{2}O use $24$ qubits, while N\textsubscript{2} uses $32$ qubits. Classical CASCI calculations provide the reference ground-state energies: $E_{\text{CASCI}}=-15.799481$~Ha for BeH\textsubscript{2}, $-76.119926$~Ha for H\textsubscript{2}O, and $-109.046672$~Ha for N\textsubscript{2}~\cite{b16}. We use the standard chemical-accuracy threshold of $1.6 \times 10^{-3}$~Ha ($1$~kcal/mol) as a visual reference in all error plots. Mean-field and CCSD calculations are performed in PySCF~\cite{b6}, producing the $t_1$ and $t_2$ amplitudes used to construct the LUCJ circuit with \texttt{ffsim}. Hardware runs use \texttt{ibmq\_fez} (IBM Heron-r2, $156$ qubits) through Qiskit Runtime's \texttt{SamplerV2}. Classical recovery and Davidson diagonalization use \texttt{qiskit-addon-sqd}. Table~\ref{tab:hparams} lists the main SQD hyperparameters.

\begin{table}[h]
\centering
\caption{SQD hyperparameters used in our experiments.}
\label{tab:hparams}
\renewcommand{\arraystretch}{1.15}
\begin{tabular}{ll}
\hline
\textbf{Parameter} & \textbf{Value} \\
\hline
Recovery iterations $T$                  & $5$ \\
Parallel batches $K$                     & $5$ \\
Samples per batch (H\textsubscript{2}O)  & $300$ \\
Samples per batch (N\textsubscript{2})   & $500$ \\
Davidson cycle                           & $300$ \\
Default shot budget                      & $10{,}000$ \\
Default layout                           & zigzag \\
Default mitigation                       & dynamical decoupling \\
                                         & + Pauli twirling \\
\hline
\end{tabular}
\end{table}

Each robustness probe varies one input while holding the others at the defaults in Table~\ref{tab:hparams}. Q1 applies three structured $t_2$ perturbations on H\textsubscript{2}O: sign flip, HOMO scaling, and zeroing; the zeroing case is also tested on BeH\textsubscript{2} and N\textsubscript{2}. Q2 crosses three N\textsubscript{2} layouts, naive at qubits $0$--$31$, randomized, and zigzag at qubits $80$--$115$, with mitigation enabled or disabled, yielding six configurations. Q3 sweeps $N_{\text{shots}} \in \{10^2,10^3,10^4,10^5\}$ on H\textsubscript{2}O and N\textsubscript{2}. We report post-recovery error relative to CASCI,
\begin{equation}
\Delta E = \left| E_{\text{SQD}} - E_{\text{CASCI}} \right|,
\label{eq:delta_e}
\end{equation}
using the final value after $T=5$ recovery iterations, with per-iteration trajectories shown in the figures.

% This document is a model and instructions for \LaTeX.
% Please observe the conference page limits. 

% \section{Ease of Use}

% \subsection{Maintaining the Integrity of the Specifications}

% The IEEEtran class file is used to format your paper and style the text. All margins, 
% column widths, line spaces, and text fonts are prescribed; please do not 
% alter them. You may note peculiarities. For example, the head margin
% measures proportionately more than is customary. This measurement 
% and others are deliberate, using specifications that anticipate your paper 
% as one part of the entire proceedings, and not as an independent document. 
% Please do not revise any of the current designations.

% \section{Prepare Your Paper Before Styling}
% Before you begin to format your paper, first write and save the content as a 
% separate text file. Complete all content and organizational editing before 
% formatting. Please note sections \ref{AA}--\ref{SCM} below for more information on 
% proofreading, spelling and grammar.

% Keep your text and graphic files separate until after the text has been 
% formatted and styled. Do not number text heads---{\LaTeX} will do that 
% for you.

% \subsection{Abbreviations and Acronyms}\label{AA}
% Define abbreviations and acronyms the first time they are used in the text, 
% even after they have been defined in the abstract. Abbreviations such as 
% IEEE, SI, MKS, CGS, ac, dc, and rms do not have to be defined. Do not use 
% abbreviations in the title or heads unless they are unavoidable.

\section{Findings}
\label{sec:findings}

% This section presents the results of the three robustness probes defined in Section~\ref{sec:approach}, ordered along the SQD pipeline from classical preprocessing to QPU sampling.

\subsection{Q1: Classical-Input Robustness}
\label{subsec:result_t2}

\begin{figure}[t]
\centering
\begin{tikzpicture}
\begin{semilogyaxis}[
  width=0.80\columnwidth,
  height=0.5\columnwidth,
  xlabel={Iteration},
  ylabel={Energy error (Ha)},
  xmin=-0.15, xmax=4.15,
  ymin=4e-4, ymax=1.5e-2,
  xtick={0,1,2,3,4},
  ytick={1e-3,1e-2},
  ymajorgrids=true,
  xmajorgrids=true,
  grid style={dashed, gray!25},
  tick label style={font=\scriptsize},
  label style={font=\scriptsize},
  legend style={
    font=\tiny,
    fill=white,
    fill opacity=0.85,
    draw opacity=1,
    text opacity=1,
    at={(0.98,0.98)},
    anchor=north east,
    row sep=-2pt,
    inner sep=1.5pt,
  },
  legend cell align={left},
  mark size=1.3pt,
  line width=0.65pt,
]
\addplot[color=blue!75!black, mark=*]
  coordinates {
    (0, 0.001113) (1, 0.002901) (2, 0.001841)
    (3, 0.001097) (4, 0.002103)
  };
\addlegendentry{clean}
\addplot[color=orange!90!black, mark=*]
  coordinates {
    (0, 0.001444) (1, 0.001077) (2, 0.000962)
    (3, 0.000616) (4, 0.001059)
  };
\addlegendentry{$\mathcal{P}_1$ sign}
\addplot[color=green!55!black, mark=*]
  coordinates {
    (0, 0.001171) (1, 0.001195) (2, 0.001306)
    (3, 0.001648) (4, 0.001263)
  };
\addlegendentry{$\mathcal{P}_2$ HOMO}
\addplot[color=red!85!black, mark=*]
  coordinates {
    (0, 0.007417) (1, 0.005340) (2, 0.002889)
    (3, 0.005963) (4, 0.004646)
  };
\addlegendentry{$\mathcal{P}_3$ zero}
\addplot[color=brown!75!black, dashed, line width=0.8pt, no marks, domain=-0.15:4.15]
  {1e-3};
\addlegendentry{chem. acc.}
\end{semilogyaxis}
\end{tikzpicture}
% \vspace{-0.5em}
\caption{H\textsubscript{2}O SQD error under structured $t_2$ perturbations.}
\label{fig:t2_attacks}
% \vspace{-0.5em}
\end{figure}
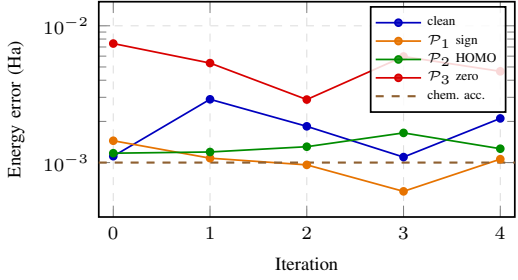

We first probe sensitivity to perturbations of the classical CCSD $t_2$ amplitudes that initialize the LUCJ ansatz. Fig.~\ref{fig:t2_attacks} shows H\textsubscript{2}O energy error per recovery iteration for the clean baseline and the three perturbations from Section~\ref{subsec:probe_t2}. The sign-flip ($\mathcal{P}_1$) and HOMO-scaling ($\mathcal{P}_2$) cases are largely absorbed, with final-iteration errors of $1.06$ and $1.26 \times 10^{-3}$~Ha, compared with the clean baseline of $2.10 \times 10^{-3}$~Ha. The strongest perturbation, $\mathcal{P}_3$, zeros all $t_2$ amplitudes; even then, recovery reduces the error from $7.42 \times 10^{-3}$~Ha at iteration $0$ to $4.65 \times 10^{-3}$~Ha at iteration $4$, about $2.2\times$ the clean floor. Table~\ref{tab:t2_zero_summary} extends the $\mathcal{P}_3$ comparison across molecules. In every tested configuration, zeroing increases the converged error by less than $7$~mHa over the clean baseline. The effect follows molecular correlation more than qubit count: BeH\textsubscript{2}, which is predominantly single-reference, changes by only $0.05$~mHa; H\textsubscript{2}O shows a $2.6\times$ degradation at $300$ samples per batch; and N\textsubscript{2}, the most strongly correlated case, shows the largest absolute clean error but only a $1.4\times$ relative degradation. For H\textsubscript{2}O, increasing samples per batch from $300$ to $500$ more than halves both clean and attack errors ($2.94 \to 1.10$ and $7.42 \to 2.89$~mHa), consistent with the shot-budget behavior discussed in Section~\ref{subsec:shot_findings}.

These results suggest that, for the perturbations tested here, the LUCJ initialization mainly affects the iteration-$0$ starting distribution. Once recovery begins, the working subspace is rebuilt from the Hamiltonian and orbital occupancies extracted from QPU samples, so perturbations that preserve support on Hartree--Fock-like configurations can be absorbed by subspace diagonalization. The cross-molecule trend in Table~\ref{tab:t2_zero_summary} is consistent with this mechanism: zeroing $t_2$ has little effect for BeH\textsubscript{2}, but a larger effect for molecules requiring more correlation. Complete zeroing is also structurally visible at circuit-construction time because it removes double-excitation terms and substantially reduces gate depth. Thus, within the tested cases, SQD is relatively insensitive to moderate $t_2$ perturbations, while severe structural corruption remains detectable before hardware execution.

\begin{table}[t]
\centering
\caption{Clean vs.\ $t_2$-zeroing attack ($\mathcal{P}_3$): SQD energy error in mHa across three molecules.}
\label{tab:t2_zero_summary}
\renewcommand{\arraystretch}{1.15}
\begin{tabular}{lccc}
\hline
\textbf{Molecule} & \textbf{Samples/batch} & \textbf{Clean} & \textbf{Attack ($\mathcal{P}_3$)} \\
\hline
BeH\textsubscript{2} ($24$ q) & $300$ & $0.00$ & $0.05$ \\
H\textsubscript{2}O ($24$ q) & $300$ & $2.94$ & $7.42$ \\
H\textsubscript{2}O ($24$ q) & $500$ & $1.10$ & $2.89$ \\
N\textsubscript{2} ($32$ q)  & $500$ & $17.39$ & $23.94$ \\
\hline
\end{tabular}
\end{table}

% The reason for this resilience is that the LUCJ ansatz's only essential role is to provide a useful iteration-$0$ starting point near the ground-state subspace; once the recovery loop runs, the subspace is rebuilt from the Hamiltonian and the average orbital occupancies extracted from QPU samples, with the LUCJ parameters playing no further role. Any perturbation that preserves QPU output support on Hartree--Fock-like configurations, which all three perturbations tested do, is therefore absorbed by the diagonalization step, and the cross-molecule trend in Table~\ref{tab:t2_zero_summary} fits this picture: when the molecule is close to its Hartree--Fock reference (BeH\textsubscript{2}), zeroing $t_2$ removes very little useful information and the recovery loop has correspondingly less work to do. $\mathcal{P}_3$ is the only perturbation that meaningfully degrades the recovered energy because it eliminates all double-excitation terms from the LUCJ circuit, but this degradation also makes it trivially detectable: the resulting circuit has substantially reduced gate depth and no two-electron excitation gates, both flaggable at construction time. CCSD amplitudes can therefore be reused from external sources without strong integrity validation, since subtle tampering is absorbed by the recovery loop while structural tampering is detectable before the circuit is ever executed.

\subsection{Q2: Layout and Mitigation Robustness}
\label{subsec:result_layout}

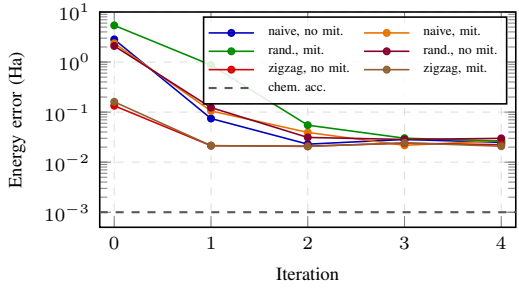
\begin{figure}[t]
\centering
\begin{tikzpicture}
\begin{semilogyaxis}[
  width=0.80\columnwidth,
  height=0.50\columnwidth,
  xlabel={Iteration},
  ylabel={Energy error (Ha)},
  xmin=-0.15, xmax=4.15,
  ymin=5e-4, ymax=1.0e1,
  xtick={0,1,2,3,4},
  ytick={1e-3,1e-2,1e-1,1e0,1e1},
  ymajorgrids=true,
  xmajorgrids=true,
  grid style={dashed, gray!25},
  tick label style={font=\scriptsize},
  label style={font=\scriptsize},
  legend style={
  font=\tiny,
  fill=white,
  fill opacity=0.88,
  draw opacity=1,
  text opacity=1,
  at={(0.98,0.98)},
  anchor=north east,
  cells={anchor=west},
  legend columns=2,
  column sep=3pt,
  row sep=-2pt,
  inner sep=1.5pt,
},
legend cell align={left},
  mark size=1.2pt,
  line width=0.6pt,
]
\addplot[color=blue!75!black, mark=*]
  coordinates {
    (0, 2.834153) (1, 0.074408) (2, 0.022888)
    (3, 0.028380) (4, 0.024874)
  };
\addlegendentry{naive, no mit.}

\addplot[color=orange!90!black, mark=*]
  coordinates {
    (0, 2.368221) (1, 0.105474) (2, 0.039264)
    (3, 0.021676) (4, 0.026919)
  };
\addlegendentry{naive, mit.}

\addplot[color=green!55!black, mark=*]
  coordinates {
    (0, 5.397600) (1, 0.872175) (2, 0.054832)
    (3, 0.030120) (4, 0.026406)
  };
\addlegendentry{rand., mit.}

\addplot[color=purple!70!black, mark=*]
  coordinates {
    (0, 2.093927) (1, 0.123174) (2, 0.031189)
    (3, 0.028423) (4, 0.029898)
  };
\addlegendentry{rand., no mit.}

\addplot[color=red!85!black, mark=*]
  coordinates {
    (0, 0.133949) (1, 0.021329) (2, 0.020870)
    (3, 0.023937) (4, 0.022318)
  };
\addlegendentry{zigzag, no mit.}

\addplot[color=brown!75!black, mark=*]
  coordinates {
    (0, 0.160488) (1, 0.021371) (2, 0.020604)
    (3, 0.024296) (4, 0.020826)
  };
\addlegendentry{zigzag, mit.}

\addplot[color=gray!70!black, dashed, line width=0.8pt, no marks, domain=-0.15:4.15]
  {1e-3};
\addlegendentry{chem. acc.}
\end{semilogyaxis}
\end{tikzpicture}
\caption{N\textsubscript{2} SQD error across layout and mitigation settings.}
\label{fig:layout_n2}
\end{figure}

We next compare three qubit layouts on \texttt{ibmq\_fez} (naive, randomized, and zigzag), each with hardware noise mitigation (DD + Pauli twirling) enabled or disabled. Fig.~\ref{fig:layout_n2} shows N\textsubscript{2} post-recovery error across the six configurations. At iteration $0$, layout dominates: zigzag, which uses bridge qubits for opposite-spin density-density interactions, gives errors of $0.13$--$0.16$~Ha regardless of mitigation, about $15$--$40\times$ lower than the naive and randomized layouts ($2.1$--$5.4$~Ha). The recovery loop rapidly collapses this spread: by iteration $2$, all configurations lie within $2.0$--$5.5 \times 10^{-2}$~Ha, and by iteration $4$ they tighten to $2.1$--$3.0 \times 10^{-2}$~Ha, reducing a $40\times$ initial gap to $1.4\times$. Within this converged band, mitigation has no consistent direction: at iteration $4$, it reduces zigzag error from $2.23$ to $2.08 \times 10^{-2}$~Ha and randomized error from $2.99$ to $2.64 \times 10^{-2}$~Ha, but increases naive error from $2.49$ to $2.69 \times 10^{-2}$~Ha. Since all three shifts are below $\sim$$3$~mHa and we do not have repeated hardware trials, we do not interpret their sign. Overall, zigzag is a useful default because it provides a much cleaner iteration-$0$ distribution, but once recovery runs, layout and mitigation differences narrow substantially.

\subsection{Q3: Shot-Budget Robustness}
\label{subsec:shot_findings}

\begin{figure}[t]
\centering
\pgfplotsset{
  every axis/.append style={
    tick label style={font=\scriptsize},
    label style={font=\scriptsize},
    title style={font=\footnotesize},
    mark size=1.4pt,
    line width=0.7pt,
  },
}
\begin{tikzpicture}
\begin{groupplot}[
  group style={group size=2 by 1, horizontal sep=0.95cm},
  width=0.52\columnwidth,
  height=0.50\columnwidth,
  ymode=log,
  xlabel={Iteration},
  xmin=-0.15, xmax=4.15,
  xtick={0,1,2,3,4},
  ymajorgrids=true, xmajorgrids=true,
  grid style={dashed, gray!25},
  legend columns=3,
  legend to name=shotlegend,
  legend style={font=\scriptsize, draw=none, fill=none,
    /tikz/every even column/.append style={column sep=6pt}},
  legend cell align={left},
]
% ---- (a) N2 ----
\nextgroupplot[title={(a) N\textsubscript{2}},
  ylabel={Energy error (Ha)},
  ymin=5e-4, ymax=1.5e1, ytick={1e-3,1e-2,1e-1,1e0,1e1}]
\addplot[color=blue!75!black, mark=*]
  coordinates {(0,6.527684)(1,0.198408)(2,0.096722)(3,0.150684)(4,0.163037)};
\addlegendentry{100 shots}
\addplot[color=orange!90!black, mark=*]
  coordinates {(0,0.201706)(1,0.024412)(2,0.020602)(3,0.030383)(4,0.016767)};
\addlegendentry{1{,}000 shots}
\addplot[color=green!55!black, mark=*]
  coordinates {(0,0.132649)(1,0.023369)(2,0.021852)(3,0.021194)(4,0.019270)};
\addlegendentry{10{,}000 shots}
\addplot[color=red!85!black, mark=*]
  coordinates {(0,0.046902)(1,0.019464)(2,0.020125)(3,0.021042)(4,0.022424)};
\addlegendentry{100{,}000 shots}
\addplot[color=brown!75!black, dashed, line width=0.9pt, no marks, domain=-0.15:4.15] {1e-3};
\addlegendentry{chemical accuracy}
% ---- (b) H2O ----
\nextgroupplot[title={(b) H\textsubscript{2}O},
  ymin=5e-4, ymax=2e-1, ytick={1e-3,1e-2,1e-1}]
\addplot[color=blue!75!black, mark=*]
  coordinates {(0,0.129931)(1,0.019239)(2,0.035413)(3,0.024681)(4,0.028282)};
\addplot[color=orange!90!black, mark=*]
  coordinates {(0,0.106567)(1,0.001645)(2,0.001497)(3,0.001242)(4,0.001506)};
\addplot[color=green!55!black, mark=*]
  coordinates {(0,0.008145)(1,0.001801)(2,0.001495)(3,0.001635)(4,0.001111)};
\addplot[color=red!85!black, mark=*]
  coordinates {(0,0.004865)(1,0.002923)(2,0.002977)(3,0.004132)(4,0.003703)};
\addplot[color=brown!75!black, dashed, line width=0.9pt, no marks, domain=-0.15:4.15] {1e-3};
\end{groupplot}
% shared legend, centered just below both panels
\node[anchor=north] at ([yshift=-2pt]current bounding box.south)
  {\pgfplotslegendfromname{shotlegend}};
\end{tikzpicture}
\caption{SQD energy error vs.\ recovery-loop iteration on \texttt{ibmq\_fez}
for N\textsubscript{2} (a) and H\textsubscript{2}O (b) at four shot budgets.}
\label{fig:shot_budget}
\end{figure}
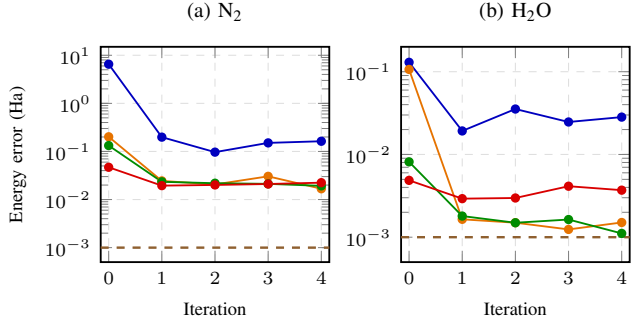

We finally examine how the recovery loop responds to QPU shot budget. Fig.~\ref{fig:shot_budget} reports post-recovery energy error per iteration for N\textsubscript{2} and H\textsubscript{2}O across four shot budgets. At iteration $0$, the $100{,}000$-shot configuration gives the lowest error for both molecules ($4.7 \times 10^{-2}$~Ha for N\textsubscript{2}, $4.9 \times 10^{-3}$~Ha for H\textsubscript{2}O), as expected from a larger raw sample. After recovery, however, the ordering changes: the $1{,}000$ and $10{,}000$-shot runs reach the lowest converged errors ($1.7$ and $1.93 \times 10^{-2}$~Ha for N\textsubscript{2}; $1.1 \times 10^{-3}$~Ha for H\textsubscript{2}O at $10^4$ shots), while the $100{,}000$-shot run plateaus at a slightly worse floor ($2.2 \times 10^{-2}$ and $3.7 \times 10^{-3}$~Ha). The $100$-shot case remains too sparse for effective recovery. This non-monotonic behavior suggests that additional shots do not necessarily enlarge the useful recovered subspace: \texttt{samples\_per\_batch} fixes the retained working set ($300$ for H\textsubscript{2}O, $500$ for N\textsubscript{2}), so increasing $N_{\text{shots}}$ changes which bitstrings enter diagonalization rather than increasing the subspace size. Thus, in our experiments, $10^3$--$10^4$ shots achieve comparable or better converged accuracy than $10^5$ shots at substantially lower QPU cost.

\section{Conclusion}
\label{sec:conclusion}

We presented an empirical robustness analysis of the SQD recovery loop on \texttt{ibmq\_fez} across three deployment choices: classical coupled-cluster singles and doubles (CCSD) amplitudes, qubit layout and noise mitigation, and QPU shot budget. The recovery loop absorbs the tested perturbations with only modest degradation in recovered energy. CCSD-amplitude perturbations remain close to the clean baseline, layout and mitigation differences narrow after a few recovery iterations, and accuracy saturates at moderate shot budgets. These results show that SQD provides practical robustness to several deployment choices, while very large shot budgets do not necessarily improve recovered accuracy.

\end{document}